\documentclass[12pt,preprint]{aastex}
\newcommand{\kms}{km\,s$^{-1}$}

\newcommand{\masyr}{mas\,yr$^{-1}$}
\newcommand{\kmspc}{km\,s$^{-1}$\,pc$^{-1}$}
\newcommand{\mura}{$\mu_{\alpha *}$}
\newcommand{\mudec}{$\mu_{\delta}$}
\newcommand{\muups}{$\mu_{\upsilon}$}
\newcommand{\mutau}{$\mu_{\tau}$}

\newcommand{\rv}{$v_{rad}$}
\newcommand{\stot}{$v$}

\newcommand{\mjup}{M$_{Jup}$}
\newcommand{\vdisp}{$\sigma_v$}
\newcommand{\mk}{M$_{K}$}
\newcommand{\parallax}{$\varpi$}
\slugcomment{accepted to {\it Astrophysical Journal}, 18 July 2005}

\shorttitle{Distance to 2M1207 B}
\shortauthors{Mamajek}

\begin{document}
\title{A Moving Cluster Distance to the Exoplanet 2M1207 B\\
in the TW Hya Association}

\author{Eric E. Mamajek\footnote{Clay Postdoctoral Fellow}}
\affil{Harvard-Smithsonian Center for Astrophysics,\\
60 Garden St., MS-42, Cambridge, MA 02138}
\email{emamajek@cfa.harvard.edu}

\begin{abstract}
A candidate extrasolar planet companion to the young brown dwarf 
2MASSW J1207334-393254 (2M1207) was recently discovered by Chauvin et al. 
They find that 2M1207 B's temperature and luminosity 
are consistent with being a young, $\sim$5 M$_{Jup}$ planet. 
The 2M1207 system is purported to be a member
of the TW Hya association (TWA), and situated $\sim$70\,pc away.
Using a revised space motion vector for TWA, and improved proper motion 
for 2M1207, I use the moving cluster method to estimate the distance to 
the 2M1207 system and other TWA members.
The derived distance for 2M1207 (53\,$\pm$\,6\,pc) forces
the brown dwarf and planet to be half as luminous as previously 
thought. The inferred masses for 2M 1207 A and B decrease to 
$\sim$21\,M$_{Jup}$ and $\sim$3-4\,M$_{Jup}$, respectively,
with the mass of B being well below the observed tip of the planetary mass
function and the theoretical deuterium-burning limit.
After removing probable Lower Centaurus-Crux (LCC) members from the
TWA sample, as well as the probable non-member TWA 22, the remaining
TWA members are found to have distances of 
49\,$\pm$\,3\,(s.e.m.)\,$\pm$\,12\,(1$\sigma$)\,pc, and
an internal 1D velocity dispersion of 0.8$^{+0.3}_{-0.2}$\,km\,s$^{-1}$.
There is weak evidence that the TWA is expanding, and the
data are consistent with a lower limit on the expansion 
age of 10\,Myr (95\% confidence).
\end{abstract}


\section{INTRODUCTION \label{intro}}
Indirect detection methods for finding extrasolar planets
have yielded in excess of 150 candidate
planets over the past decade \citep{Marcy05}. None have been 
directly imaged, however two have had their light detected through 
transit studies with the {\it Spitzer Space Telescope}
\citep[HD 209458b and TRes-1b; ][]{Charbonneau05,Deming05}. 
Recently, an object has been imaged, and resolved, whose 
properties appear somewhat consistent with being a young extrasolar 
planet: the companion to 2MASSW J1207334-393254 
\citep[2M1207; ][]{Chauvin04}. 
These objects appear to represent the 
opening chapter in humanity's quest to study the atmospheres of
planets beyond our solar system. 

As one of the nearest and youngest brown dwarfs yet identified, 
2M1207 has received considerable attention since its discovery. 
\citet{Gizis02} discovered 2M1207 in a 
spectroscopic survey of red 2MASS sources, and  
claimed that the object was a $\sim$10 Myr-old,
$\sim$25\,\mjup\, member of the nearby ($\sim$55 pc) 
TW Hya association \citep[TWA;][]{Webb99}. Further observations of
its radial velocity \citep{Mohanty03} and proper motion 
\citep{Scholz05} are roughly consistent with TWA membership. 
With low resolution spectroscopy, 
\citet{Gizis02} found 2M1207 to
show signs of low surface gravity and
strong H$\alpha$ emission (EW\,=\,300\,$\AA$).
In their echelle spectroscopy survey, \citet{Mohanty03} 
found the H$\alpha$ emission line to be broad and asymmetric, 
and accompanied by several other Balmer and \ion{He}{1} emission
lines. \citet{Mohanty03} hypothesize that the brown dwarf is probably
still accreting from a circumstellar disk.

While astrophysically interesting in its own right as
a representative of the new class of young, accreting brown dwarfs
\citep[e.g.][]{Muzerolle05,Mohanty05}, it appears that 2M1207 may 
become most famous for being the host ``Sun'' for the first 
imaged extrasolar planet -- if indeed 2M1207B can be called a ``planet''.
\citet{Chauvin04} discovered a faint companion
to 2M1207, which has near-IR photometry and a low signal-to-noise-ratio 
spectrum consistent with having a late-L spectral type.  
Recently, \citet{Chauvin05} and Schneider et al. (in prep.) 
confirmed that the companion B is indeed 
co-moving with 2M1207 A. Debate on the origin and classification 
of this object is in its infancy. To help better constrain 
the physical nature of
this object, I present an improved distance estimate
to the 2M1207 system through the moving cluster method. 
The new distance provides more accurate luminosities 
(and inferred masses) for the components of this
interesting substellar binary.

\section{ANALYSIS \label{analysis}}

Although a trigonometric parallax is not yet available for
2M1207, one can exploit the star's putative membership in the
TW Hya association to derive a distance using the cluster
parallax (or ``moving-cluster'') method 
\citep[e.g.][]{Atanasijevic71,deBruijne99a}.
With an observed proper motion and radial velocity
(as well as other supporting evidence), I test whether
the star is consistent with being a TWA group member. 
To exploit this technique, one needs to take the following steps: 
(1) estimate the space motion vector for the TWA, 
(2) test whether the observations for 2M1207 
(proper motion, radial velocity) are consistent with the TWA 
motion vector, and
(3) use the moving cluster method to estimate the parallax
from the proper motion and TWA space motion data. I address
these steps in order. Although the rest of the TWA membership
is not the focus of this study, I will briefly mention
relevant results for these systems throughout this analysis. 
I will also examine whether the expansion of the TWA is
detectable, and whether it can help constrain the age of 2M1207
and the rest of the association. 

\subsection{Sample \label{sample}}

The initial pool of candidate TWA members considered in
this study are listed in Table \ref{tab:TWA}. 
I add to TWA numbers 1 through 25 \citep{Zuckerman04} 
the three, new, low-mass candidate members
2M1207 and 2MASSW J1139511-315921 from \citet{Gizis02}, and 
SSSPM J1102-3431 from \citet{Scholz05}. 
The TWA members from \citet[][TWA 1-11]{Webb99} and 
\citet[][TWA 12, 13]{Sterzik99} comprise what I will
tentatively call the ``classic'' membership of the TW Hya association. 
These are young stars which were mostly selected due
to infrared or X-ray excesses within the immediate vicinity
of TW Hya. 
There has been debate
regarding the membership for TWA 14-19
\citep{Mamajek01,Lawson05}, since their positions, 
proper motions, and rotational properties
are at variance with TWA 1-13. 
\citet{Mamajek01} and \citet{Lawson05} have
suggested that TWA 14-19 are probably
members of the more distant \citep[$\sim$120\,pc; ][]{deZeeuw99} 
and older \citep[$\sim$16 Myr; ][]{Mamajek02}
Lower Centaurus Crux OB association (LCC). 
TWA 20 was claimed to be a TWA member by \citet[][R03]{Reid03}, but rejected
by \citet{Zuckerman04} due to its weak Li. As the Li data is not
published, and the similarity in proper motion between TWA 20 and
the other TWA members is quite striking, I retain TWA 20 in the candidate
pool.
TWA 21 through 25 were selected by
\citet[][SZB03]{Song03} due to their strong Li and H$\alpha$ emission.
From Fig. 6 of SZB03, it appears that 
TWA 23 and 25 have positions and proper motions very close to
those of TWA 1-13,
but TWA 21 and 22 are spatially isolated, and
TWA 24 has a small proper motion, similar to LCC members. Hence it is not
obvious that many of the TWA 14-25 stars were born in the same
star-formation event as TWA 1-13. I conservatively
include only the classic members (TWA 1-13) in 
the initial calculations for estimating the convergent 
point and space motion vector for the TW Hya association. 

\subsection{Astrometric Data \label{data}}

The adopted proper motion and radial velocity
data for proposed TWA members, and their associated references, 
are presented in Table 1. 
I searched the literature and on-line catalogs\footnote{
ADS (http://adsabs.harvard.edu/),
Vizier (http://vizier.u-strasbg.fr/viz-bin/VizieR), and
SIMBAD (http://simbad.harvard.edu/sim-fid.pl)} 
to find the best values 
for the proper motions and radial velocities of TWA members. 
To mitigate against the effects of short-term astrometric perturbations
by short-period companions, I preferentially adopted
the long-baseline proper motion with the smallest error
bars \citep[usually Tycho-2 or UCAC2; ][]{Hog00,Zacharias04} over
{\it Hipparcos} values \citep{Perryman97}, when available. In a few instances,
I calculated new proper motions using published positions. 
I calculated weighted
mean radial velocities when multiple values were available,
or adopted systemic velocities for spectroscopic binaries,
when available. 

2M1207 has two published proper motion estimates
in the literature \citep{Gizis02,Scholz05}. The
\citet{Gizis02} proper motion (\mura,\,\mudec\,=\,
--100,\,--30\,\masyr)
does not have error bars and is based only on 
a few plate images in the USNO image archive. 
\citet{Scholz05} estimated a proper motion for 2M1207 of 
\mura\,=\,--78\,$\pm$\,11\,\masyr, \mudec\,=\,--24\,$\pm$\,9\, \masyr.
This proper motion estimate included a Chandra
pointing, rather than an actual measured
position, and so is invalid. Omitting the Chandra pointing, 
R.-D. Scholz has calculated 
a revised proper motion of \mura\,=\,--67\,$\pm$\,7\,\masyr, 
\mudec\,=\,--28\,$\pm$\,11\, \masyr\,
using a least-squares fit with equal weighting
(R.-D. Scholz, personal communication). 

\begin{figure}
\epsscale{1.0}
\plotone{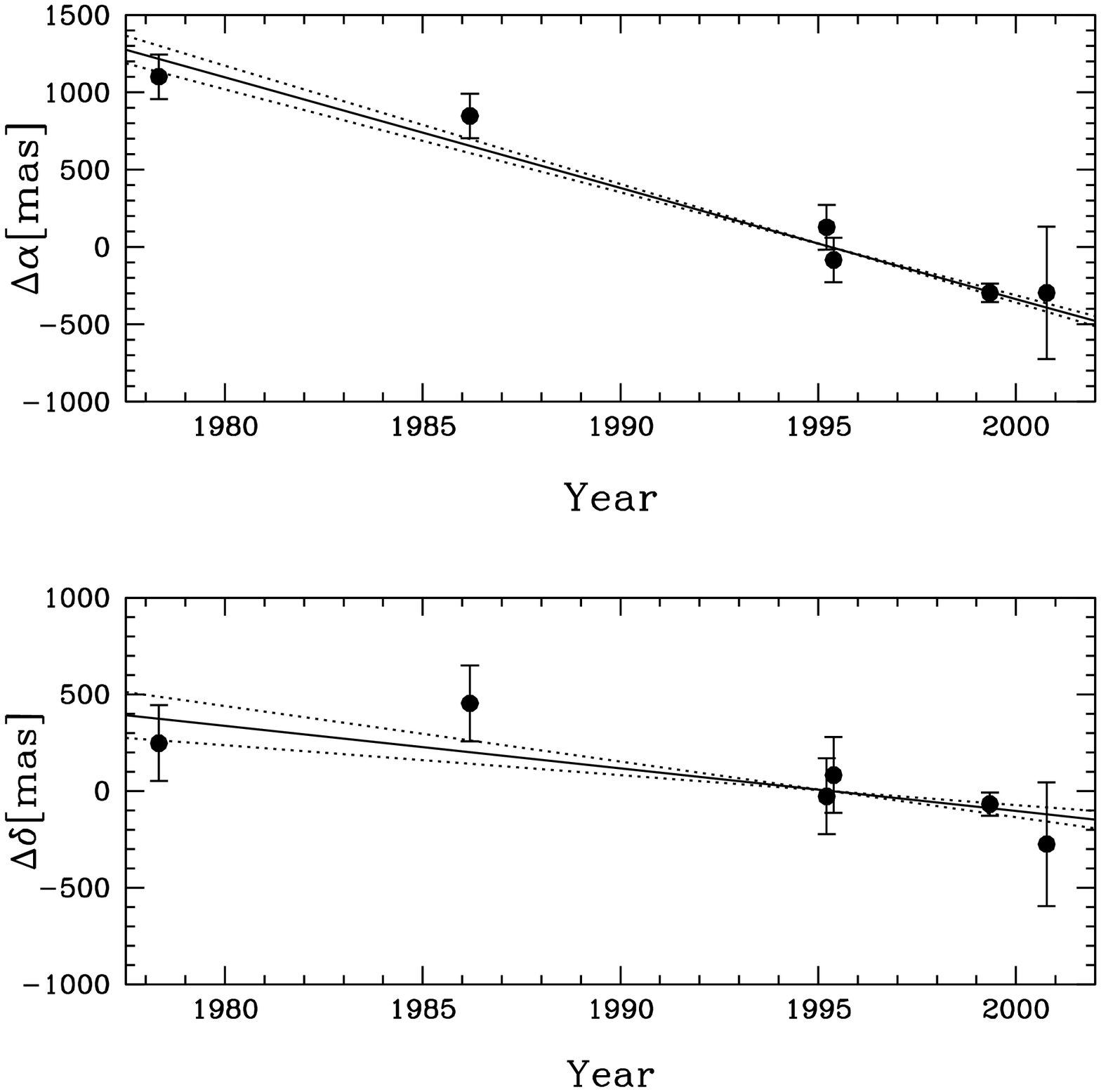}
\caption{\label{fig:pm}
The relative change in position of 2M1207 since the late 1970s
in $\alpha$ ({\it top}) and $\delta$ ({\it bottom}).
The position offsets are relative to the positions on the ICRS at mean epoch. 
The mean epoch positions and mean epochs are $\alpha_o$ = 181$^{\circ}$.889563
($t_o$($\alpha_o$) = 1995.08) and $\delta_o$ = --39$^{\circ}$.548315
($t_o$($\delta_o$) = 1996.57). The effects of parallax were not
included since the predicted amplitude ($\sim$20\,mas) is smaller than 
the individual position errors. The positional errors at mean
epoch are $\sigma_{\alpha_o}$ $\simeq$ $\sigma_{\delta_o}$ $\simeq$ 50\,mas.
The data are consistent with \mura\,=\,$-71.6\,\pm\,6.7$\,\masyr\, and 
\mudec\,=\,$-22.1\,\pm\,8.5$\,\masyr\, for 2M1207.}
\end{figure}

As there are large differences
in the accuracy between the SuperCOSMOS, 2MASS, and DENIS positions
($\sim$60\,mas vs. $\sim$500\,mas), I recalculated
the proper motion using weighting by the inverse of the square of the
positional errors, following the method of 
\citet{Corbin77}\footnote{The formulae are given in the on-line 
documentation for the AC2000.2 catalog \citep{Urban98} at
http://ad.usno.navy.mil/ac/}.
The SuperCOSMOS and 2MASS positions are tied to the
International Celestial Reference System (ICRS) via the
Tycho-2 catalog, and so for our purposes they are on the same
system.
In order to estimate a positional error for the
SuperCOSMOS positions, I performed a least-squares fit to the 4 SuperCOSMOS
points, and found their scatter consistent with positional errors
of $\sigma_{\alpha *}$ = 143 mas and $\sigma_{\delta}$ = 196 mas.
These errors are very consistent with the SuperCOSMOS positional 
errors quoted by \citet{Hambly01}.
I corrected the DENIS position for the 2MASS-DENIS offset found by
\citet{Cabrera-Lavers03}, since the 2MASS positional
errors are much smaller than DENIS's \citep[2MASS is tied to 
the ICRS via the Tycho-2 catalog to an accuracy of 
$\sim$80 mas; ][]{Cutri03}.
For the DENIS positional errors, I adopted the square root
of the 2MASS-DENIS rms differences added in quadrature with the
2MASS-ICRS rms residuals ($\sim$80\,mas), giving 
$\sigma_{\alpha *}$ = 430\,mas, and $\sigma_{\delta}$ = 320\,mas.
I estimate the proper motion of 2M1207 to 
be {\mura\,=\,--72\,$\pm$\,7\,\masyr}, {\mudec\,=\,--22\,$\pm$\,9\, \masyr\,},
which is within 1$\sigma$ of both of Scholz's estimates.
The change in the position of 2M1207 over time is plotted in Fig. \ref{fig:pm}.

\subsection{The Space Motion of the TWA \label{spacemotion}}

In order to calculate a cluster parallax for 2M1207, we
require an accurate convergent point solution for the TW Hya 
association, to which 2M1207 is proposed to be a member.
Mean space motion vectors and/or convergent point solutions
for the TWA were previously estimated by 
\citet{Frink01}, \citet[][MF01]{Makarov01}, R03, and SZB03. 
Considering the increase
in proposed association membership (SZB03), wealth
of new proper motion data \citep[UCAC2;][]{Zacharias04} and
radial velocity data \citep{Torres03} made available
since R03, I will briefly discuss and
reanalyze the kinematics of TWA.
To estimate the space motion for the TWA, I
will combine information from 2 different methods:
using what little data there is regarding the 3D space
motion vectors for individual members, as well as applying
the convergent point method on the classical membership. 

Only four classic TWA members have sufficient data to reliably 
calculate the 3D space motion vector (TWA 1, 4, 9, and 11), 
and these individual determinations have modest 
errors in any given velocity component
\citep[$\sigma$ $\sim$ 1-2\,\kms; ][]{Mamajek00}. Three of
the systems are binaries, but their systemic velocities
are probably accurate to $\sim$1\,\kms, or better. The 
mean barycentric Galactic space motion vector for these four
systems ($U, V, W$ = --10.2, --17.1, --5.1\,\kms) provides the
best estimate of the {\it centroid} velocity vector for the
TW Hya association. 

For helping refine the vertex estimate for the
TWA, I will use the
convergent point method on the classical membership. 
The convergent point, as calculated only from the proper
motion data, will also become important when the question
of association expansion is addressed (\S\ref{expansion}).
I approximately follow the convergent point grid 
technique of \citet{Jones71}.
In this implementation, I alter Jones's definition of $t^2$, following
\citet{deBruijne99a}, and include an intrinsic velocity dispersion term 
(\vdisp\,=\,1\,\kms), and assumed distance (50\,pc) in the definition
for $t^2$ (the method is rather insensitive to both input values). 
Over the entire hemisphere ${\alpha\,\in\,0^{\circ}-180^{\circ}}$, 
I calculate the $t^2$ statistic at every 0$^{\circ}$.1 grid step, and
find the celestial position which gives the minimum $t^2$ value. 
For every grid point, the method assumes that this
position is the convergent point for the group, and rotates the 
stellar proper motion components (in \mura\, and \mudec) 
to the proper motion directed toward the convergent point
(\muups) and perpendicular to the great circle joining
the star and test convergent point (\mutau). The method
iteratively searches for which test convergent point 
minimizes the ${\tau}$ components of proper motion 
for the input sample.
Jones's and de Bruijne's $t^2$ value can be treated statistically 
as the classic ${\chi}^2$ \citep{Bevington92}.  
In its iterative search for the group convergent point,
the method will reject stars contributing the
most to the $t^2$ statistic, until the position with lowest $t^2$ value 
corresponds to a sufficiently high enough $\chi^2$ probability that the best
convergent point can not be statistically rejected.  
For a statistical rejection threshold, I adopt a 5\% level of 
significance (i.e. 5\% probability of falsely rejecting the null hypothesis) 
following \citet{Trumpler53}.  

The TWA stars are sufficiently convergent, and the proper
motion errors for the faint members are large enough, 
that a convergent point can be determined for all the classic 
TWA members (\#1-13) 
with a low $\chi^2_{\nu}$ ($\chi^2$/$\nu$ = 15.9/13; 
$\chi^2$ probability = 25\%). For internal velocity
dispersions of \vdisp\,$>$\,0.6\,\kms, the method is able to 
find a convergent point with $\chi^2$ probability 
of $>$5\% without rejecting any of the classic members. 
If \vdisp\,=\,0.6\,\kms\, is adopted,
TWA 6 (which contributes the most to the $t^2$ statistic)
is rejected, and a sound solution is found with the other
nuclear members ($\chi^2$ probability = 21\%). 
The internal velocity dispersion is probably near
\vdisp\,$\simeq$\,1\,\kms, and with this adopted 
velocity dispersion, 33\% of the classical members 
contribute $\Delta$$t^2$ $>$ 1 (similar to how
MF01 estimate the velocity dispersion). Hence,
there is no good reason to remove TWA 6 in the hunt for
a statistically satisfactory convergent point for TWA 1-13. 
I will determine a more refined estimate of the velocity
dispersion for TWA in \S\ref{distance}. 
The ability of the technique to give a statistically sound
convergent point solution ($\chi^2_{\nu}$ $\simeq$
1) with \vdisp\,=\,1\,\kms\, already suggests that
the velocity dispersion of TWA is similar to that
of nearby OB associations \citep{Madsen02}.

\begin{figure}
\epsscale{1.0}
\plotone{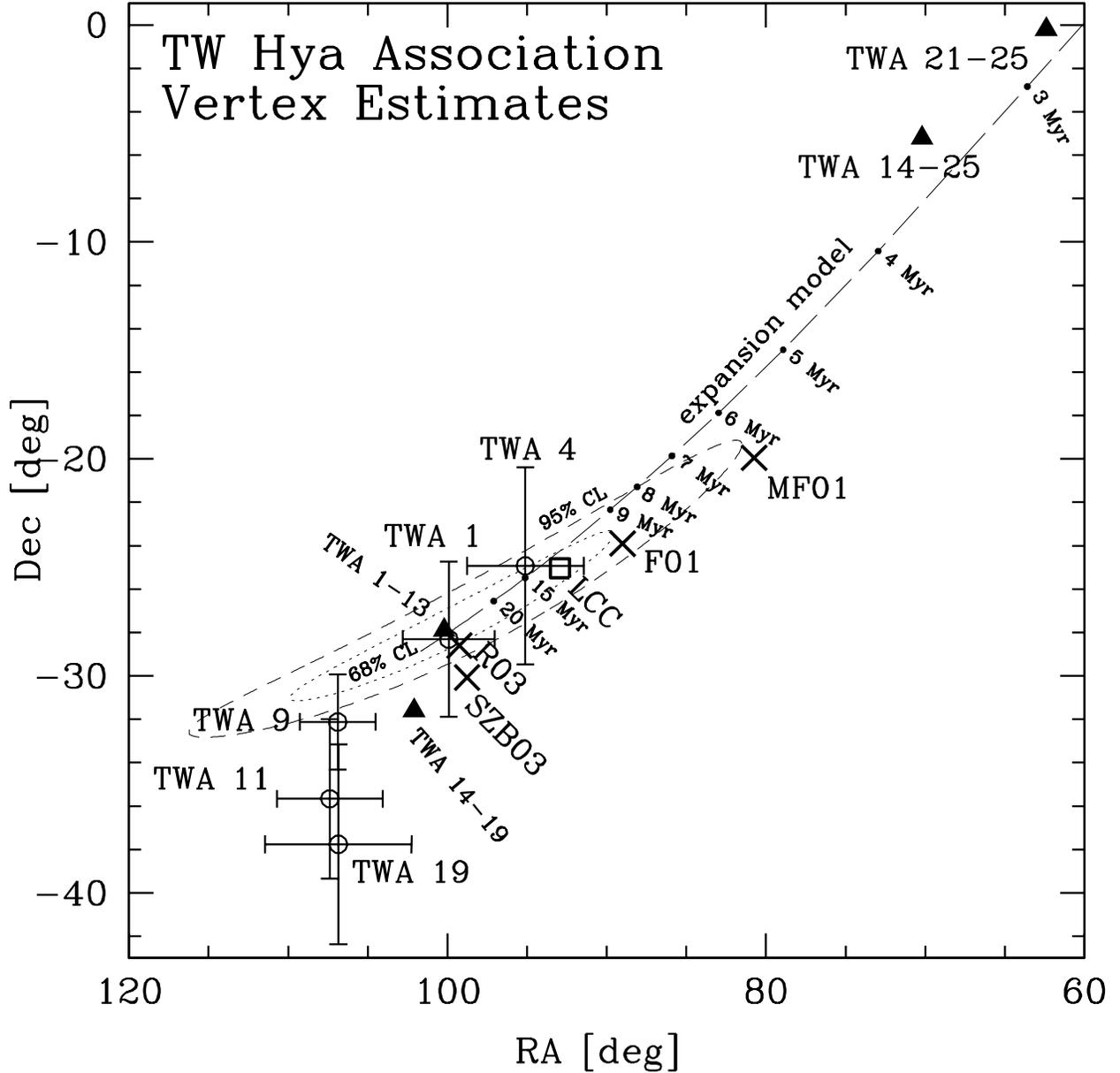}
\caption{\label{fig:cvp}
New and previously determined convergent point (vertex) estimates 
for the TW Hya association. The {\it open circles} are
the inferred convergent points for TWA 1, 4, 9, 11, and 19 based on 
their UVW space motions (and 1$\sigma$ error bars). {\it Large Xs} are previously published 
TWA vertices from \citet[][F01]{Frink01}, MF01, R03, and SZB03. 
{\it Filled triangles} are
the vertices for several subsamples of TWA objects 
(TWA 1-13, 14-19, 21-25, and 14-25) found through the convergent 
point method (based solely on the proper motion data).
The {\it dotted line} and {\it dashed line}
are the 68.3\% (1$\sigma$) and 95.5\% (2$\sigma$) confidence levels
in $\alpha$ and $\delta$ around the vertex found for the ``classic'' 
TWA sample (TWA 1-13). The error regions around the other filled triangles (TWA 14-25,
21-25, 14-19) are 50-100\% larger than that for TWA 1-13, and similarly shaped
(but not shown for clarity). Predicted vertices for the TWA with 
a wide range of expansion ages are distributed along the {\it long dashed 
line}.}
\end{figure}

In Fig. \ref{fig:cvp}, I plot the convergent points for subsamples of 
the TW Hya association, as well as previous determinations from
the literature. 
I also plot the convergent points for subsamples of the TWA 14-25 membership
in Fig. \ref{fig:cvp}. The confidence regions of these subsamples
are roughly twice as large as than that for TWA 1-13, but not wholly 
inconsistent given the large error bars. Much of the positional deviance 
of these subsample convergent points is owed to TWA 22, which may not
be a kinematic TWA member (\S\ref{distance}). 
From Fig. \ref{fig:cvp}, one can conclude that the 
convergent point for TWA 1-13 (within the dashed confidence
regions) 
agrees well with that predicted by the TWA space motion vectors
found by R03 and SZB03. The TWA vertex found by MF01 is just outside
of the 95\% confidence region, and seems to be deviant when compared
to the values from R03, SZB03, F01, and the results of this convergent
point analysis. 
This fact, combined with the finding that most of the stars in
the MF01 convergent point analysis are not pre-MS \citep{Song02},
suggests that their convergent point and dynamical age for the TWA (8.3 Myr) are 
not valid. I will discuss the expansion age further in \S\ref{expansion}.

After considering the
agreement between the TWA 1-13 convergent point and
that inferred from the mean TWA space motions from
R03 and SZB03,
I adopted the following fiducial TWA parameters.
For the group vertex, I took the weighted mean of
the vertices from my convergent point analysis of TWA 1-13
($\alpha, \delta$ = 100$^{\circ}$\,$\pm$\,10$^{\circ}$, 
{--28$^{\circ}$\,$\pm$\,4$^{\circ}$}), and the individual vertices
inferred from the space motion vectors for TWA 1, 4, 9, and
11 \citep[using eqn. 10 of ][]{deBruijne99a}. This analysis
assumes zero association expansion, and assumes that there is
no significant offset between spectroscopic and physical
radial velocities -- both of which are acceptable assumptions
at this level of accuracy. The best estimate of the 
convergent point for the TWA is calculated to be 
($\alpha$\,=\,{103$^{\circ}$.2\,$\pm$\,1$^{\circ}$.5}, 
$\delta$\,=\,{--30$^{\circ}$.7\,$\pm$\,1$^{\circ}$.5}). 
For the mean speed of the classic TWA membership, 
I adopt the weighted mean barycentric speed 
($v$\,=\,21.3\,$\pm$\,1.3\,(s.e.m.)\,\kms) of TWA 1, 4, 9, and 11, 
using their astrometry and radial velocities in Table \ref{tab:TWA}, and
weighted mean Hipparcos and Tycho parallaxes. 

\subsection{Is 2M1207 a TWA Member? \label{member}}

Given the adopted convergent point solution for the ``classic'' TWA members,
and a proper motion for 2M1207, one can estimate a membership probability
and predict the star's radial velocity. Using the
updated proper motion for 2M1207 (\S\ref{data}), I find that
most of the motion is indeed pointed towards the convergent point
(\muups\,=\,75\,$\pm$\,7\,\masyr) and very little of it is 
in the perpendicular direction (\mutau\,=\,2\,$\pm$\,8\,\masyr). 
Using the membership probability equation from \citet[][
his eqn. 23]{deBruijne99a}, and adopting a mean cluster
distance of 50\,pc and velocity dispersion of 1\,\kms, I estimate
a membership probability of 98\%. This membership
probability should be interpreted as: given the proper motion
errors, 98\% of bona fide TWA members are expected to have \mutau\, values 
more deviant than 2M1207. 
That is, the proper motion of 2M1207 
is consistent with the null hypothesis (\mutau\,=\,0) for an 
``ideal'' member. One can also use the predicted and observed radial velocity
as a check of the moving cluster method. 
Assuming parallel motion among group members, the method predicts 
the radial velocity as \rv\,=\,\stot\,cos$\lambda$, where \stot\, is
the speed of the group, and $\lambda$ is the angular separation
(62$^{\circ}$.9\,$\pm$\,1$^{\circ}$.5) between 2M1207
and the convergent point. 
The predicted radial velocity for 2M1207 
(+9.7\,$\pm$\,1.6\,\kms) is within 0.6$\sigma$ of the
observed radial velocity measured by 
\citet[][+11.2\,$\pm$\,2\,\kms]{Mohanty03}. Both
the proper motion and radial velocity data for 2M1207
are quantitatively consistent with TWA membership, and
its evidence of membership is as strong as that for most of
the classical members.

\subsection{Distances \label{distance}}

\subsubsection{The Distance to 2M1207}

If a star belongs to a moving group, its proper
motion can be used to estimate its distance.
The star's moving cluster parallax (\parallax) is calculated as 
\parallax\,=\,$A$\,\muups/\stot\,sin\,$\lambda$, where \muups, \stot, 
and $\lambda$ are as described before,
and $A$ (= 4.74047) is the astronomical unit expressed in 
the convenient units of km\,yr\,s$^{-1}$ \citep{deBruijne99a}. Using the
values (and uncertainties) for \muups, \stot, and $\lambda$
as given in \S\ref{data} and \S\ref{spacemotion}, 
I calculate a cluster parallax for
2M1207 of \parallax\,=\,18.8\,$\pm$\,2.3\,mas, or
a corresponding distance of $d$\,=\,53\,$\pm$\,6 pc. 
The only published distance estimates to 2M1207
are $\sim$70\,pc \citep[][]{Chauvin04}
and 70\,$\pm$\,20\,pc \citep[][]{Chauvin05}. Both
are photometric distance estimates which
force 2M1207A to be an unreddened M8 star on a 10 Myr-old 
isochrone. Considering the variations between
published evolutionary tracks, especially for stars which
are young and low-mass \citep{Hillenbrand04}, the cluster
parallax distance should be considered an improved estimate.

\subsubsection{Distances to TWA Objects: Implications and Final Membership}

The agreement between the trigonometric parallaxes for TWA 1, 4, 9, and
11, and their cluster parallaxes are excellent, as shown in 
Table \ref{tab:par}. All the parallaxes are within 2$\sigma$
of each other, with an insignificant weighted-mean zero-point offset
of --0.8\,$\pm$\,1.2\,mas, in the sense ``cluster minus
trigonometric''. Cluster parallax distances for all TWA member
candidates are given in column 9 of Table \ref{tab:TWA}. 

\begin{figure}
\epsscale{1.0}
\plotone{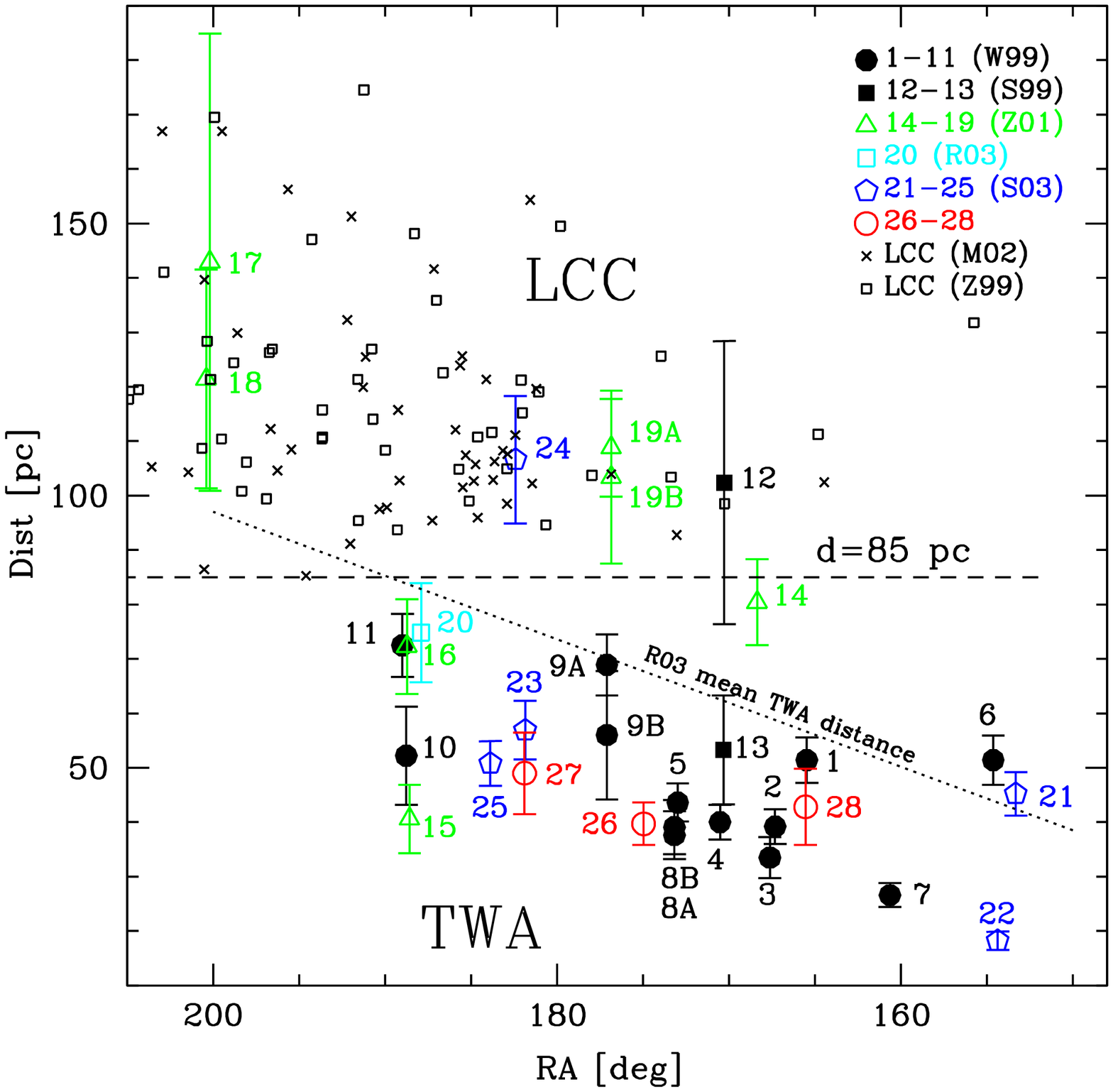}
\caption{\label{fig:ra_dist}
Right Ascension versus distance for candidate members of TWA and LCC.
{\it Solid black circles} are TWA 1-11 from \citet{Webb99}, 
{\it solid black squares} are TWA 12-13 from \citet{Sterzik99},
{\it open green triangles} are TWA 14-19 from \citet{Zuckerman01},
{\it open cyan square} is TWA 20 from \citet{Reid03},
{\it open blue pentagons} are TWA 21-25 from \citet{Song03},
{\it open red circles} are TWA 26-28 (2M1139, 2M1207, SSSPM J1102),
{\it small Xs} are pre-MS LCC members from \citet{Mamajek02},
and {\it small open squares} are B-type LCC members from \citet{deZeeuw99}.
The {\it dotted line} is the mean TWA distance relation
from \citet{Reid03}. The {\it dashed line} at $d$ = 85\,pc is
shown to illustrate the detached nature of the TWA and LCC groups,
i.e. although they have similar space motions, they do
appear to occupy separate regions. 
If the likely LCC members (TWA 12, 17, 18, 19,and 24) are excluded,
as well as the probable nonmember TWA 22, the rest of the TWA membership is 
consistent with having distances of $d$\,=\,49\,$\pm$\,12 (1$\sigma$) pc.}
\end{figure}

There is a small caveat regarding the cluster parallax distances in 
Table \ref{tab:TWA} and Fig. \ref{fig:ra_dist} that is worth
elaborating upon. 
There have been suggestions that some TWA stars may actually
be background members of the 
$\sim$16-Myr-old Lower Centaurus-Crux (LCC) OB subgroup
at $d$ $\simeq$ 110\,pc \citep[e.g.][]{Mamajek01,Mamajek02}. 
The space motion vectors for TWA and LCC are very similar,
and within roughly $\sim$5\,\kms\, \citep{Mamajek00}. 
If one calculates cluster parallax 
distances to ``TWA'' objects using the space motion vector of LCC 
\citep{Madsen02}, the mean distances in column 7 of Table 1 change 
by less than $\pm$5\% (rms). This is smaller than the quoted distance
errors (typically $\sim$11\%). Hence, any conclusions based on the 
distribution of cluster parallax distances (i.e. Fig. \ref{fig:ra_dist})
are very insensitive to whether individual ``TWA'' objects are co-moving
with either TWA or LCC.

I plot the cluster parallax
distances versus Right Ascension in Fig. \ref{fig:ra_dist}. 
Fig. \ref{fig:ra_dist} illustrates that there appears to be
a gap in the distances between LCC members and ``classic''
TWA members near $d$\,=\,85\,pc, effectively splitting the groups spatially.
Hence, TWA 12, 17, 18, 19, and 24 have distances 
more consistent with LCC than the other TWA members.
Previous investigators (MF01, R03) have suggested that the
TWA members are clustered at distances of $\sim$70\,pc, however
Fig. \ref{fig:ra_dist} suggests that what is really being seen
is two detached populations of young stars: one at 
$\sim$50\,pc (TWA) and one at $\sim$110\,pc (LCC). 
The agreement between the observed and predicted radial velocities
for TWA 12, 17, 18, 19, and 24, are probably due to the similarity in
space motion between LCC and TWA (see Fig. \ref{fig:cvp}). 
As it is often not clear how ``TWA'' candidate members have been
retained (or rejected) in past studies, 
there may be an observational bias present for the radial velocities
of these more distant objects to agree well with that of the 
foreground members. 

As Fig. \ref{fig:ra_dist} suggests that some of the ``TWA'' stars may be
more distant members of LCC, it is worth reexamining the vertex
of the remaining TWA members, including the new brown dwarf 
members TWA 26-28. 
If the convergent point method (\S\ref{spacemotion}) is run on the
remaining members (again assuming a mean distance of 50\,pc and
\vdisp\,=\,1\,\kms), a somewhat poor vertex solution
is found ($\chi^2$/$\nu$ = 39.8/23; $\chi^2$ probability = 1.6\%).
The biggest contributer to the $\chi^2$ (contributing a third
of the quantity) is the closest TWA candidate -- TWA 22.
If TWA 22 is dropped, the convergent point method shifts
by a few $\sigma$ in position, and a much more statistically sound vertex
is found: 
$\alpha\,=\,100^{\circ}.5\,\pm\,5^{\circ}.0$, 
$\delta\,=\,-27^{\circ}.9\,\pm\,2^{\circ}.3$, 
$\chi^2$/$\nu$ = 17.5/22; $\chi^2$ probability = 74\%. Rejecting
further members has negligible effect on the vertex, and only pushes
the $\chi^2$ probability to absurdly higher levels. This
remarkable reduction in $\chi^2$, upon removal of TWA 22 from the
sample, suggests that TWA 22 should probably be excluded as a TWA member. 
Clearly it is a nearby, young star, however it does not
appear to be a kinematic TWA member. 
In the initial calculation of membership probabilities (column 8
of Table \ref{tab:TWA}; \S\ref{member}), TWA 22 had $P$\,=\,2\% -- by
far the lowest. This new {\it a posteriori} convergent point 
estimate is currently the best that can be done purely geometrically, i.e. with 
proper motions alone. It is in excellent agreement
with the original TWA 1-13 vertex determination, and with the
individual vertices for TWA 1, 4, 9, and 11. 

With the sample of ``final'' TWA members (denoted ``Y'' or ``Y?'' in column 11 
of Table 1), one can independently estimate the velocity dispersion $\sigma_v$ of
TWA based on
how well the proper motions determine the convergent point. Considering
the range of $\chi^2$ values for an acceptable fit \citep[see discussion
in ][]{Gould03}, the final estimate of the velocity dispersion of
TWA, from the proper motion data alone, 
is \vdisp\,=\,$0.8^{+0.3}_{-0.2}$\,\kms.
By adopting \vdisp\,=\,0.8\,\kms, the uncertainties on the proper 
motion-determined convergent point decrease to $\sigma_{\alpha}$\,=\,4$^{\circ}$.2
and $\sigma_{\delta}$\,=\,1$^{\circ}$.9. 
Using the revised, proper motion-based convergent point estimate,
and the new estimate of the velocity dispersion of the group,
has negligible effect on the distance determinations. 
For these reasons (and clarity of presentation),
I have chosen not to list the reevaluated quantities.

After excluding TWA 12, 17, 18, 19, 22, and 24 as TWA members, I characterize
the final TWA membership with 
probability plots \citep[so as to be immune to the effects
of outliers;][]{Lutz80}:
$d_{TWA}$ = 49\,$\pm$\,3\,(s.e.m.)\,$\pm$\,12\,(1$\sigma$)\,pc, 
$\alpha_{TWA}$ = 174$^{\circ}$.8\,$\pm$\,12$^{\circ}$.3\,(1$\sigma$),
and $\delta_{TWA}$ = --37$^{\circ}$.1\,$\pm$\,7$^{\circ}$.5\,(1$\sigma$).
The projected radii in $\alpha$ and $\delta$ correspond to $\sim$7\,pc at 
$d$ = 49\,pc. Taking into account the typical distance errors ($\sim$5\,pc)
and the observed distance dispersion ($\sim$12\,pc), the data are
consistent with the radius along the line of sight being $\sim$10\,pc
($\sim$40\% larger than the projected width of $\sim$7\,pc). 
All three of the new
brown dwarf members (TWA 26-28; red open circles in Fig. \ref{fig:ra_dist}) 
cluster lie between $d$\,=\,40-53\,pc, close to the classic
TWA membership.

With the membership and characteristics of the TWA better defined,
one can ask the question: are there other stars in the vicinity whose
astrometric data also suggest that they are TWA members? Dozens of other
young, low-mass, field stars have been proposed as TWA members, enough so that
assessing their membership is probably worth a separate study. A question
that can be answered here is: {\it are there any high-mass
TWA members besides HR 4796?}. The quoted magnitude limits of the 
{\it Hipparcos} catalog suggest that it should be complete for unreddened 
A and B-type stars on, or above, the main sequence within $\sim$85\,pc
\citep{Perryman97}. 
I queried the {\it Hipparcos}
database for stars within a 15$^{\circ}$ radius centered on the
TWA central position given earlier. I retained the 31 stars with
parallaxes of $>$10\,mas and $B-V$ colors of $<$0.30 (consistent with
unreddened stars earlier than F0). I calculated membership probabilities
and predicted cluster parallaxes for these stars, in the same manner
that was done for 2M1207 in \S\ref{member}. Of these 31 stars, only eight had
membership probabilities of $>$5\%. For these eight, I compared the
moving cluster parallax values to the {\it Hipparcos} trigonometric
parallaxes. Only three of these eight stars had agreement between cluster
and trigonometric parallaxes at better than 2$\sigma$: HR 4796 (known
member), HIP 54477 (A1V, $d$\,=\,58\,pc), and HIP 53484 (F0V, $d$\,=\,97\,pc).
HIP 53484 is $\sim$4$\sigma$ more distant than the mean TWA distance, and
nearly $\sim$15$^{\circ}$ from the TWA centroid position, so
I reject its TWA membership, and discuss it no further. HIP 54477 
is not so easy to dismiss
as a TWA member. This A1 dwarf has a high TWA membership probability (90\%), and its
trigonometric parallax (17.2\,$\pm$\,0.7\,mas) agrees fairly 
well with its predicted TWA
cluster parallax (20.8\,$\pm$\,1.6\,mas). Its projected position is in the core
region near TWA 2, 4, and 8. At $d$ $\simeq$ 56\,pc ({\it Hipparcos}) or
$d$ $\simeq$ 48\,pc (predicted cluster parallax distance), it would be
slightly further than TWA 2, 4, and 8 (all of which have 
$d$ $\simeq$ 40\,pc). The radial velocity of HIP 54477 
is not well constrained \citep[\rv\,=\,+16.2\,$\pm$\,10\,\kms;][]{Barbier-Brossat00}, 
but consistent
with that for a TWA member at its position (+12.2\,$\pm$\,1.6\,\kms). 
The star appears to be close to the zero-age main sequence, and
so could be as young as the other TWA members. Further observations
should be undertaken to see if the object has any low-mass companions
which may further constrain its age. The membership of HIP 54477 to TWA 
can not be rejected on kinematic grounds, but one would certainly like to see
further data before claiming that it is a true TWA member.  
In summary, {\it the TWA appears
to contain at least one (HR 4796), but possibly two (HIP 54477), stars
hotter than F0 in its membership.}

\subsection{Expansion Age of TWA \label{expansion}}

One may be able to put an interesting astrophysical constraint on the
age of the 2M1207 system through calculating an ``expansion age'' 
for the TW Hya association.
MF01 claimed that the kinematics of the TWA are consistent
with an expansion age of 8.3 Myr. The analysis of MF01 included tens of
X-ray-selected stars in their analysis which have been since 
shown to not be pre-MS stars \citep[SZB03,][]{Torres03}. 
As the majority of the stars in the MF01 analysis are not genetically 
related to TW Hya or its cohort, {\it this expansion age is not a 
useful constraint on the age of the TWA or 2M1207.}
With the best proper motion and radial velocity data currently available, 
I investigate whether an expansion is still evident in the TWA using a 
Blaauw expansion model. 
For discussions on trying to detect the linear expansions of unbound 
associations, see \citet{Blaauw56,Blaauw64}, \citet{Bertiau58}, 
\citet{Jones71}, \citet{Brown97}, \citet{Dravins99}, and \citet{Madsen02}.

\subsubsection{The Blaauw Linear Expansion Model \label{Blaauw_model}}

Linear expansion of an association can not be demonstrated with 
proper motions alone \citep{Blaauw64,Brown97}. A group of stars
with generally parallel motion vectors, but with a small linear
expansion, will simply appear to converge to a point further away 
(higher $\lambda$) than that demonstrated by a group with strictly 
parallel motion vectors. The classical convergent point method
equations which assume parallel motion are slightly modified to
allow for expansion. In the \citet{Blaauw64} linear expansion model,
the individual cluster parallax ($\varpi$) for an association member is calculated as:
\begin{equation}
\varpi\,=\,\frac{\mu_{\upsilon}\,A\,}{\,v^{\prime}\,\sin\,\lambda^{\prime}}
\end{equation}
and the radial velocity is predicted to follow the relation:
\begin{equation}
v_{rad}\,=\,v^{\prime}\,\cos\,\lambda^\prime\,+\,\kappa\,d\,+\,K
\end{equation}
where $A$ is the AU as previously defined,
$\mu_{\upsilon}$ is the proper motion directed toward the convergent point,
$\kappa$ is the expansion term in units of \kmspc,
$d$ is the distance to the star in pc (where 
$d_{pc}$\,$\simeq$\,1000\,$\varpi_{mas}^{-1}$), 
and $K$ is a zero-point term which may reflect
gravitational redshift or convective blueshift terms \citep[see][for 
a detailed discussion]{Madsen03}. The ``expansion age'' $\tau$ of the
association in Myr is:
\begin{equation}
\tau\,=\,\gamma^{-1}\,\kappa^{-1}
\end{equation}
where $\gamma$ is the conversion factor 1.0227\,pc\,Myr$^{-1}$\,km$^{-1}$\,s. 
The cluster speed \stot$^\prime$ and star-vertex angular separation 
$\lambda^\prime$ are defined differently that in the standard case of 
parallel motions. In the Blaauw model, \stot$^\prime$ is the 
barycentric speed of a 
hypothetical 
association member participating in the expansion, situated at the 
barycenter of our solar system 
\citep[see Fig. 3 of][]{Blaauw64}, and $\lambda^\prime$ is the 
angular separation between a star and the association convergent point
as defined solely by the stars' proper motions. If an association
is expanding, the convergent point determined from the mean 3D space
motion of its members (the ``centroid'' space motion) will define
a different ``convergent point'' than the vertex determined through
a convergent point analysis of the stars' proper motions.

To test whether the association is expanding or not, and possibly
assign an ``expansion age'', I analyze the
data for TWA members two ways. First, I compare model convergent
points for varying expansion ages to the observed convergent point.
Second, I will use the available radial 
velocities and cluster parallax distances to directly measure the 
expansion rate. 

\subsubsection{Expanding versus Non-expanding Association Convergent Point 
\label{cvp_compare}}

In Fig. \ref{fig:cvp}, I plot the variation in the convergent point
(long-dashed line)
if one takes the TWA ``centroid'' space motion vector (using the mean velocity 
vector for TWA 1, 4, 9, and 11), and add linear expansion with
characteristic expansion timescales. 
In \S\ref{distance}, I determined that the best convergent point
for the final TWA membership using the proper motion data alone was
($\alpha\,=\,100^{\circ}.5\,\pm\,4^{\circ}.2$, 
$\delta\,=\,-27^{\circ}.9\,\pm\,1^{\circ}.9$).
Predicted expansion model convergent points for ages 0-100 Myr were statistically 
compared to the observed convergent point error ellipse. 
From this analysis alone, {\it one can reject expansion ages of $<$7\,Myr 
at 5\% significance, and $<$6\,Myr at 1\% significance}. 

The close agreement between the TWA vertex found by the convergent point method 
and the vertices for the four individual TWA members with known UVW vectors 
(see Fig. \ref{fig:cvp}), suggests that any kinematic expansion must be very subtle, 
and perhaps not even demonstrable with existing data. It is worth exploring
whether the radial velocity data can help either determine a significant
expansion age, or at least place a more interesting lower limit. 

\subsubsection{A ``Hubble Diagram'' for TWA?\label{Hubble}}

\citet{Blaauw56,Blaauw64} suggested that linear expansion or contraction may
be detectable if deviations are present between the observed
spectroscopic radial velocities, and those predicted from the
moving group method for parallel motion. If a significant 
linear expansion term $\kappa$ is present, then 
the Blaauw expansion model equations (Eqns. 1 and 2) predict that one 
should see a correlation between distance
$d$ and the difference (\rv\,--\,\stot\,cos$\lambda$) between the observed and predicted 
spectroscopic radial velocities.
As the radius of the TWA is $\sim$10\,pc and the isochronal age is 
$\sim$10$^7$\,yr, one expects that $\kappa$ should
be of order $\sim$0.1\,\kmspc, if the stars are linearly
expanding from a point.  

The effects of expansion on cluster parallax
distances are usually negligible \citep[e.g. as shown by comparing
trigonometric parallaxes to cluster parallaxes; e.g.][]{deBruijne99b,Mamajek02,Madsen02}. 
For the case of TWA, the change in cluster parallax distances, between 
assuming parallel
motion and linear expansion, is $<$6\% rms for expansion ages of $>$5\,Myr,
and $<$3\% rms for $>$10 Myr. Note that expansion ages of $<$6\,Myr were effectively
ruled out in \S\ref{cvp_compare}, and the typical distance errors from other sources 
(e.g. proper motions) are $\sim$11\%. One can then conclude that the effects 
of association expansion (if any) on the distances and distance errors 
quoted in this study are negligible.

In order to detect any possible expansion by fitting the Blaauw model
to the observations, I adopt the convergent point defined solely
using the proper motion data.
I estimate $v^{\prime}$ for the four TWA members (TWA 1, 4, 9, 11) 
with trigonometric parallaxes through the equation:
\begin{equation}
v^{\prime}\,=\,\frac{\mu_{\upsilon}\,A\,}{\,\varpi\,\sin\,\lambda^{\prime}}
\end{equation}
The mean value for the four TWA members is $v^{\prime}$ = 20.4\,$\pm$\,2.2\,\kms. 
Already, one notices that $v$ (21.3\,$\pm$\,1.3\,\kms) is indistinguishable
from $v^{\prime}$, which is consistent with no expansion. 

\begin{figure}
\epsscale{1.0}
\plotone{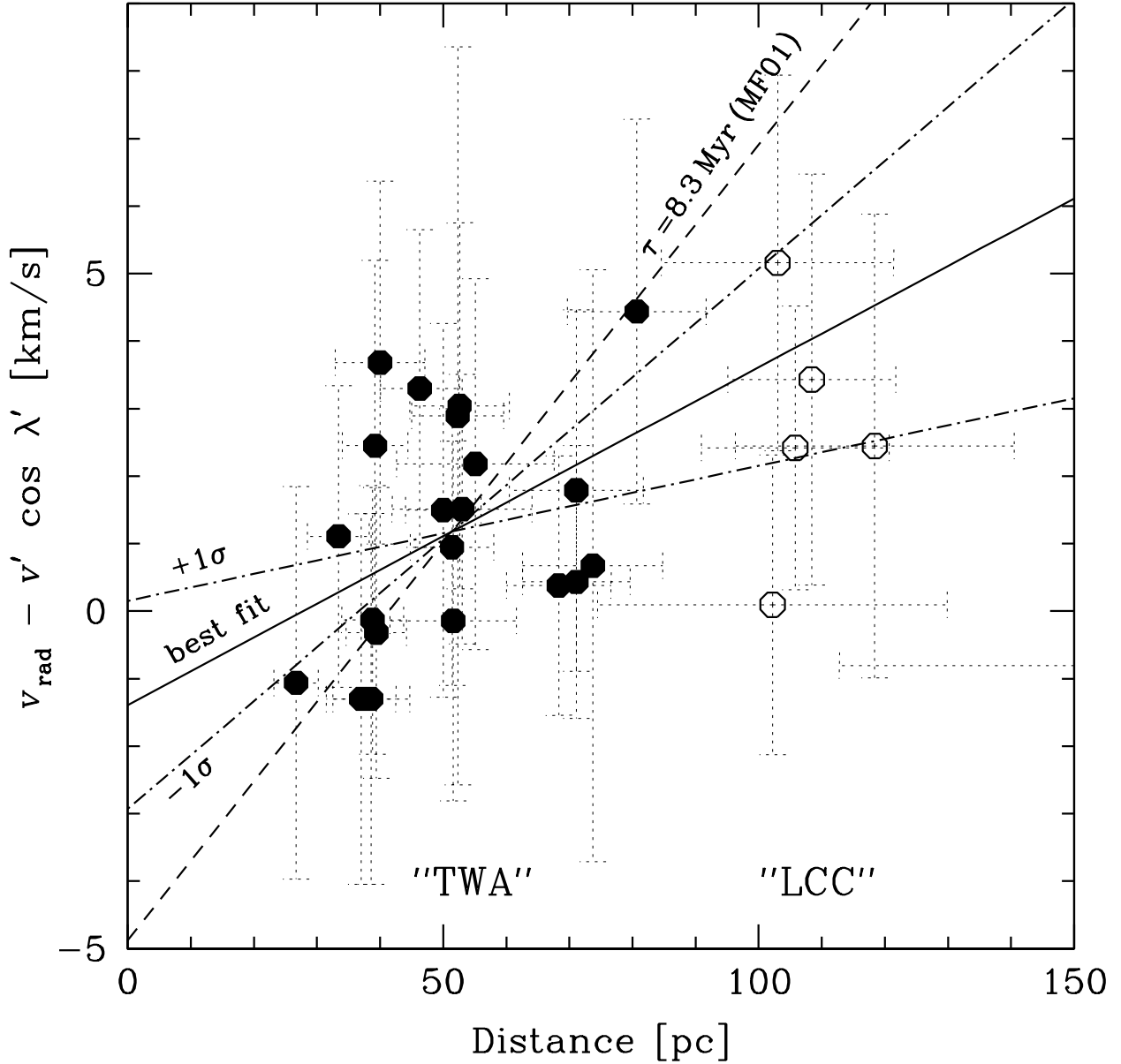}
\caption{\label{fig:rv_kappa}
Distance versus the difference between the observed radial 
velocity (\rv) and the distance-dependent radial velocity 
component of the Blaauw expansion model (\stot\,cos$\lambda$). 
``Final'' TWA members are filled circles,
and probable LCC members are open circles. For parallel
motion and no expansion, a slope of zero is expected. 
In the case of expansion, one expects the closest members to be 
more blueshifted, and the more distant members to be more redshifted.
The {\it solid line} is the best fit slope $\kappa$ 
(= 0.049\,$\pm$\,0.027\,\kms)
to the data, with 1$\sigma$ error bars as {\it dotted-dashed lines}.
The prediction for a system with expansion age
of 8.3 Myr (MF01) is shown by the {\it dashed line}, and ruled out by
the data.}
\end{figure}

In order to see whether a non-zero $\kappa$ coefficient is detectable, I
plot in Fig. \ref{fig:rv_kappa} the data in the format $d$ versus 
($v_{rad} - v^{\prime}\,cos\lambda^{\prime}$) so that one can solve for the
slope $\kappa$ and intercept $K$:
\begin{equation}
v_{rad}\,-\,v^{\prime}\,\cos\,\lambda^\prime\,=\kappa\,d\,+\,K
\end{equation}
Plotted in this form, any expansion will manifest itself as a significantly
positive slope.
The individual distance estimates for the expansion model are calculated as:
\begin{equation}
d_{pc}\,=\,\frac{1000\,v^{\prime}\,sin\,\lambda^{\prime}}{A\,\mu_{\upsilon}}
\end{equation}
As seen in Fig. \ref{fig:rv_kappa}, it is a success of the kinematic model 
that the ($v_{rad} - 
v^{\prime}\,cos\lambda^{\prime}$) values
are crowded near zero at all. Recall that the {\it predicted} radial velocity
component ($v^{\prime}\,cos\lambda^{\prime}$) is totally
independent of {\it any} measured radial velocity data, i.e. they are solely dependent
on the the convergent point position (via $\lambda^{\prime}$), and the 
trigonometric parallax distances and proper motions for TWA 1, 4, 9, and 11 
(via $v^{\prime}$). This agreement further strengthens the interpretation that the 
TWA constitutes a bona fide kinematic group.  

The errors in distance and velocity difference have some peculiarities worth
mentioning. The distance errors tend to 
scale with distance, i.e. $\sigma_d$\,$\propto$\,$d$. 
Secondly, the distances will all be affected systematically if the convergent 
point is in error. Finally, the linear fit of the data to equation \#5 
using weighting in both variables \citep[using {\it fitexy} from Numerical 
Recipes; ][]{Press92} gives an
uncomfortably good fit ($\chi^2$/$\nu$ = 7.8/19), presumably due to overestimated
errors in either the observed radial velocities, group speed, or convergent point. 
This weighted fit finds
$\kappa$\,=\,+0.036\,$\pm$\,0.039\,km\,s$^{-1}$\,pc$^{-1}$ and 
$K$\,=\,+1.07\,$\pm$\,0.51\,\kms, but again, due to the very low $\chi^2$, it is unclear
how much to believe the errors. 

To avoid overinterpreting a derived slope $\kappa$ whose error bars may
not be believable, I fit an unweighted, ordinary least-squares line
to the data, with the distance $d$ as the independent variable, and the velocity
difference ($v_{rad} - v^{\prime}\,cos\lambda^{\prime}$) as the dependent variable. 
I do this for the 19 TWA ``final'' members whose radial velocity
errors are $<$2.5\,\kms. Since the sample is small, I use bootstrap and jackknife
resampling to help determine the error in the derived slope \citep{Feigelson92},
although the agreement with the errors derived from the asymptotic formulae is 
excellent. The least-squares fit finds $\kappa$\,=\,+0.049\,$\pm$\,0.027\,km\,s$^{-1}$\,pc$^{-1}$
and $K$\,=\,+1.20\,$\pm$\,0.36\,\kms\, (evaluated at the mean distance). 
Although the sign of the slope is consistent with expansion, 
the correlation is very weak (Pearson $r$ = 0.42\,$\pm$\,0.19). 
The basic result is unchanged
whether all of the TWA members are retained, independent of radial velocity
error, or if only the 8 TWA stars with radial velocity errors of $<$2\,\kms\, are 
retained\footnote{A non-zero velocity offset $K$ should not cause too much alarm.
Part of the offset may be due to gravitational redshift, which for the typical
$\sim$10\,Myr-old TWA member with mass $\sim$0.5\,M$_{\odot}$ should be of
order $\sim$0.4\,\kms \citep[][using radii from the D'Antona \& 
Mazzitelli 1997 tracks]{Greenstein67}, compared to $\sim$0.6\,\kms\, for the Sun. 
An unexplained radial velocity offset of 0.4\,\kms\, appears to be present 
among low-mass Hyades members \citep{Gunn88}, even after accounting for 
gravitational redshift. The offsets between measured ``spectroscopic'' 
radial velocities and ``astrometric'' radial velocities are difficult to
quantify, but should be more easily measurable for larger samples of stars 
with future astrometric missions \citep{Dravins99,Madsen03}.}.

Although the slope $\kappa$ is small, one can state that it is positive
at 95\% confidence, i.e. that the data are consistent with some expansion.
Unfortunately, the derived expansion age has very large errors, and
is of limited utility:
$\tau\,=\,\gamma^{-1}\,\kappa^{-1}$\,$\simeq$\,20$^{+25}_{-7}$\,Myr. 
The probability distribution function of $\kappa$ {\it excludes} 
expansion ages of $<$8.7\,Myr at 99\% confidence,
and $<$10.4\,Myr at 95\% confidence. The confidence intervals on
the expansion age are very wide: 13-43 Myr (68\% CL) and 9.5-262 Myr (90\% CL), 
with $\sim$4\% of the probability distribution corresponding to contraction.
The expansion age advocated by MF01 (8.3\,Myr) can, however, be statistically
rejected. 
{\it It does not seem appropriate at this time to quote an 
unambiguous ``expansion age'' for the TW Hya association, but to quote
the lower limit ($\gtrsim$10.4\,Myr)}. 

\section{DISCUSSION \label{discussion}}

With an improved distance estimate, one can revise the absolute
magnitude, luminosity, and inferred mass estimates for 2M1207A and B. 
The properties of 2M1207 A and B from the literature, and derived here, 
are listed in Table \ref{tab:phot}.
Using the photometry from \citet{Chauvin04} and revised distance
estimate from the moving cluster method, the absolute magnitudes of 
2M1207A and B are \mk(A)\,=\,8.32\,$\pm$\,0.27 and \mk(B)\,=\,13.30\,$\pm$\,0.29 mag,
respectively. These are 0.6 mag fainter than one would derive using
$d$ = 70\,pc (i.e. a factor of two intrinsically dimmer). I calculate
luminosities using these absolute magnitudes, and the bolometric 
correction estimates of \citet{Golimowski04}.  
Using the constraints 
on luminosity and age \citep{Chauvin05},
I interpolate masses from the non-gray evolutionary tracks of 
\citet{Burrows97}, the DUSTY tracks of \citet{Chabrier00}, and the
COND tracks of \citet{Baraffe03}. For all three sets of evolutionary
tracks, the masses of A and B cluster near $\sim$21\,\mjup\,
and $\sim$3-4\,\mjup. Table \ref{tab:phot} also lists the mass extrema
from the 1$\sigma$ extrema in both luminosity and age (i.e. the low 
mass end is for the -1$\sigma$ luminosity {\it and} age, and the high mass 
end is the +1$\sigma$ luminosity {\it and} age). With the
previous distance estimates ($\sim$70\,pc), \citet{Chauvin04} estimates
mass of 25\,\mjup\, and 5\,\mjup\, for A and B, respectively. 
For all three models, the inferred
upper mass limit of 2M1207 B ($\sim$5-7\,\mjup) is less than half of the 
deuterium-burning mass limit \citep[$\sim$13\,\mjup;][]{Burrows97},
and less than half of the maximum mass of Doppler velocity planets
\citep[$\sim$15\,\mjup;][]{Marcy05}. Hence, 2M1207 B could be 
considered a ``planet'' on the merits of its inferred mass. 

The angular separation 
of AB (778 mas) measured by \citet{Chauvin04} translates into a 
projected physical separation of 41\,$\pm$\,5 AU at the revised distance 
(similar
to the semi-major axis of Pluto). If the observed separation is
assumed to be equal to the semi-major axis, and one adopts masses of 
3.5 and 21\,\mjup\, for A and B, then one naively predicts an orbital 
period of $\sim$1700\,yr. The pair has a high mass ratio
($q$ $\sim$ 0.2), and B is massive enough to
force the primary to be $\sim$6\,AU from the system barycenter.  
A solid detection of orbital evolution 
(or any hint of the dynamical masses of the components) will 
probably not be reported anytime soon.

One can not rule out whether the TWA
is expanding on a timescale longer than its isochronal age ($>$10\,Myr).  
The slow, or negligible, expansion may also be a clue that the proto-TWA molecular
cloud complex was perhaps not a small pc-sized core with tens of stars,
similar to those seen in Taurus 
(e.g. LDN 1551). The TWA members may 
have formed in a series of small-N systems (N $\sim$ few stars) distributed 
along filaments, separated by a few pc, and with similar bulk motions. 
The TWA appears to be moving away from the LCC 
subgroup \citep{Mamajek00}, so it is conceivable that the proto-TWA cloudlets
were simply fragments of the proto-LCC cloud, which owed their
velocities to molecular cloud turbulence \citep{Feigelson96}.
An alternative scenario is that the proto-TWA cloudlets were
bright-rim clouds or cometary globules on the periphery of 
LCC $\sim$10-15 Myr ago, when presumably the LCC subgroup still had a 
few late-O stars \citep{deGeus92}. Such cloudlets could have
been accelerated away from the LCC O-star population through
the rocket effect \citep{Oort55}, and compressed to form 
stars due to radiation-driven implosion \citep{Bertoldi90}. 
The energy input from deceased 
LCC members (via UV light, winds, and supernovae) 
has probably dominated the energy input of the local interstellar 
medium over the past 10 Myr, 
and within 100\,pc, in the general direction of LCC and TWA 
\citep{Maiz-Apellaniz01}. Small-scale star-formation
in cometary globules on the edge of OB associations has strong
observational support \citep{Reipurth83,Ogura98},
and there is strong evidence for triggering by the massive
stellar population \citep[e.g.][]{Kim05,Lee05}.
A cometary globule formation scenario for TWA might explain a few 
observational quirks of the group, namely its location 
($\sim$70\,pc away from the nearby LCC OB subgroup),
age ($\sim$7 Myr younger than LCC), space motion vector 
\citep[directed $\sim$5\,\kms\,away from the LCC;][]{Mamajek00}, 
and low stellar density. 
The small, young stellar groups associated with $\eta$ Cha, 
$\epsilon$ Cha, and $\beta$ Pic show many
of these same symptoms \citep{Mamajek99,Mamajek00,Ortega02,Jilinski05},
although the $\eta$ and $\epsilon$ Cha clusters appear to be more strongly
bound than the TWA and $\beta$ Pic groups. 
Cloudlets analogous to those on the periphery of Vel OB2 \citep{Kim05} 
and Ori OB1 \citep{Lee05} may be the evolutionary predecessors of
small, unbound, $\sim$10 Myr-old associations like TWA.  
That 2M1207 and the TWA formed in a region of rather 
low stellar density could explain how such a wide, low-mass binary system 
as 2M1207 could survive its birth environment intact.

\acknowledgments

EM is supported by a Clay Postdoctoral Fellowship from the
Smithsonian Astrophysical Observatory. The author
thanks the referee, Ronnie Hoogerwerf, for useful comments and
criticisms which improved the paper. The author also thanks 
Glenn Schneider, Ralf-Dieter Scholz, Willie Torres, 
Jay Farihi, and Subu Mohanty for useful discussions, and
Lissa Miller and Kevin Luhman for critiquing early drafts. 
This publication makes use of data products from the Two Micron All
Sky Survey (2MASS), which is a joint project of the Univ. of Massachusetts
and the IPAC/California Institute of
Technology, funded by NASA and the NSF.
SuperCOSMOS Sky Survey material is based on photographic data originating 
from the UK, Palomar and ESO Schmidt telescopes and is provided by the 
Wide-Field Astronomy Unit, Institute for Astronomy, University of 
Edinburgh. The GSC 2.2 is a joint project of STSci and the Osservatorio 
Astronomico di Torino. STSci is operated by AURA, for NASA under contract 
NAS5-26555. The participation of the Osservatorio Astronomico di Torino is 
supported by the Italian Council for Research in Astronomy. Additional 
support for GSC 2.2 was provided by ESO, ST-ECF, the International GEMINI project 
and the ESA Astrophysics Division. This work made extensive use of the 
SIMBAD database, operated at the CDS, Strasbourg, France.


\clearpage
\begin{deluxetable}{llcccccrcll}
\tabletypesize{\scriptsize}
\setlength{\tabcolsep}{0.03in}
\tablewidth{0pt}
\tablecaption{Properties of Proposed TW Hya Association Members \label{tab:TWA}}
\tablehead{
{(1)}&{(2)}  &{(3)}     &{(4)}     &{(5)} &{(6)}  &{(7)} &{(8)}       &{(9)}      &{(10)}&{(11)}\\
{TWA}&{Other}&{\mura}   &{\mudec}  &{Ref.}&{Prob.}&{$d$} &{\rv(pred.)}&{\rv(obs.)}&{Ref.}&{Final}\\
{\#} &{Name} &{[\masyr]}&{[\masyr]}&{}    &{[\%]} &{[pc]}&{[\kms]}    &{[\kms]}   &{}    &{Member?}\\
}
\startdata 
1   &TW Hya                 & -66.8\,$\pm$\,1.6  & -15.2\,$\pm$\,1.3 & 1& 97& 51\,$\pm$\,4  &+13.2&+12.7\,$\pm$\,0.2& 7      & Y\\
2   &CD-29 8887             & -91.6\,$\pm$\,1.8  & -20.1\,$\pm$\,1.3 & 2& 57& 39\,$\pm$\,3  &+12.4&+11.0\,$\pm$\,0.1& 7      & Y\\
3   &Hen 3-600              &-109.3\,$\pm$\,8.7  &  +0.8\,$\pm$\,8.9 & 3& 18& 34\,$\pm$\,4  &+12.9&+12.5\,$\pm$\,1: & 8,9    & Y\\ 
4  &HD 98800                & -91.7\,$\pm$\,1.5  & -30.0\,$\pm$\,1.5 & 1& 73& 40\,$\pm$\,3  &+11.0& +9.3\,$\pm$\,1: & 10,11  & Y\\
5AB &CD-33 7795             & -85.4\,$\pm$\,3.6  & -23.3\,$\pm$\,3.7 & 1& 88& 44\,$\pm$\,4  &+11.4& var.         & 7      & Y\\
6   &TYC 7183-1477-1        & -57.0\,$\pm$\,2.1  & -20.9\,$\pm$\,2.1 & 1& 53& 51\,$\pm$\,5  &+15.4&+16.9\,$\pm$\,5  & 9      & Y\\
7   &TYC 7190-2111-1        &-122.2\,$\pm$\,2.3  & -29.3\,$\pm$\,2.3 & 4&100& 27\,$\pm$\,2  &+14.2&+11.7\,$\pm$\,2  & 9      & Y\\
8A  &GSC 06659-01080        & -99.3\,$\pm$\,9.0  & -31.3\,$\pm$\,8.9 & 3& 83& 38\,$\pm$\,4  &+10.5& +7.8\,$\pm$\,2  & 9      & Y\\  
8B  &2MASS J11324116-2652090& -95.3\,$\pm$\,10.0 & -29.5\,$\pm$\,10.3& 3& 84& 39\,$\pm$\,5  &+10.5& +7.8\,$\pm$\,2  & 9      & Y\\ 
9A  &CD-36 7429A            & -52.8\,$\pm$\,1.3  & -20.2\,$\pm$\,1.8 & 1& 81& 69\,$\pm$\,6  &+10.6& +9.5\,$\pm$\,0.4& 7      & Y\\
9B  &CD-36 7429B            & -70.7\,$\pm$\,13.3 &  -6.6\,$\pm$\,15.8& 3& 67& 56\,$\pm$\,12 &+10.6&+11.3\,$\pm$\,2  & 9      & Y\\ 
10  &GSC 07766-00743        & -72.6\,$\pm$\,12.2 & -32.1\,$\pm$\,12.3& 3&100& 53\,$\pm$\,9  & +8.3& +6.6\,$\pm$\,2  & 9      & Y\\
11A &HR 4976A               & -53.3\,$\pm$\,1.3  & -21.2\,$\pm$\,1.1 & 4& 99& 73\,$\pm$\,6  & +8.0& +6.9\,$\pm$\,1.0& 9,12,13& Y\\
12  &RX J1121.1-3845        & -36.3\,$\pm$\,8.6  &  -1.6\,$\pm$\,8.9 & 3& 69&103\,$\pm$\,26:&+12.4&+10.9\,$\pm$\,1.0& 7,14   & N\\
13  &RX J1121.3-3447        & -67.4\,$\pm$\,11.8 & -17.0\,$\pm$\,11.8& 3& 99& 53\,$\pm$\,10 &+12.0&+12.1\,$\pm$\,1: & 7,14   & Y\\ 
14  &UCAC2 12427553         & -43.4\,$\pm$\,2.6  &  -7.0\,$\pm$\,2.4 & 1& 96& 80\,$\pm$\,8  &+13.1&+16.0\,$\pm$\,2  & 9      & Y?\\
15  &GSC 08236-01074        &-100.0\,$\pm$\,33.0 & -16.0\,$\pm$\,6.0 & 5& 88& 41\,$\pm$\,6  & +9.1&+11.2\,$\pm$\,2  & 9      & Y\\
16  &UCAC2 12217020         & -53.3\,$\pm$\,5.2  & -19.0\,$\pm$\,5.2 & 1&100& 72\,$\pm$\,9  & +8.8& +9.0\,$\pm$\,2  & 9      & Y\\
17  &GSC 08248-00700        & -28.0\,$\pm$\,8.5  & -11.1\,$\pm$\,8.5 & 3& 84&163\,$\pm$\,46:& +6.3& +4.6\,$\pm$\,6  & 9      & N\\
18  &UCAC2 12908626         & -29.0\,$\pm$\,5.2  & -21.2\,$\pm$\,5.2 & 1& 64&121\,$\pm$\,20:& +6.0& +6.9\,$\pm$\,3  & 9      & N\\
19A &HD 102458A             & -33.6\,$\pm$\,0.9  &  -8.5\,$\pm$\,0.9 & 1& 73&109\,$\pm$\,9: &+11.6&+13.5\,$\pm$\,2.4& 7,9    & N\\
19B &HD 102458B             & -35.6\,$\pm$\,4.8  &  -7.5\,$\pm$\,4.6 & 1& 97&103\,$\pm$\,16:&+11.6&+15.2\,$\pm$\,2  & 9      & N\\
20  &GSC 08231-02642        & -52.0\,$\pm$\,5.0  & -16.0\,$\pm$\,6.0 & 5& 97& 75\,$\pm$\,9  & +9.0& +8.1\,$\pm$\,4  & 15     & Y\\  
21  &HD 298936              & -65.3\,$\pm$\,2.4  & +13.7\,$\pm$\,1.0 & 1& 99& 45\,$\pm$\,4  &+15.8&+17.5\,$\pm$\,0.8& 16     & Y\\
22  &SSSPM J1017-5354       &-176.0\,$\pm$\,7.0  & -22.0\,$\pm$\,8.0 & 6&  2& 18\,$\pm$\,2: &+15.5& $\ldots$        & $\ldots$&N?\\
23  &SSSPM J1207-3247       & -68.0\,$\pm$\,4.0  & -23.0\,$\pm$\,4.0 & 6& 86& 57\,$\pm$\,5  & +8.9& $\ldots$        & $\ldots$&Y\\
24  &MML 5                  & -34.4\,$\pm$\,2.8  & -13.1\,$\pm$\,1.7 & 1& 10&107\,$\pm$\,12:&+11.1&+11.9\,$\pm$\,0.9& 16     & N\\
25  &TYC 7760-283-1         & -75.0\,$\pm$\,2.0  & -26.9\,$\pm$\,1.4 & 1&100& 51\,$\pm$\,4  & +9.2& +9.2\,$\pm$\,2.1& 16     & Y\\
26  &2MASSW J1139511-315921 & -93.0\,$\pm$\,5.0  & -31.0\,$\pm$\,10.0& 6& 99& 40\,$\pm$\,4  &+10.6&+11.6\,$\pm$\,2  & 17     & Y\\
27  &2MASSW J1207334-393254 & -71.6\,$\pm$\,6.7  & -22.1\,$\pm$\,8.5 & 3& 98& 53\,$\pm$\,6  & +9.7&+11.2\,$\pm$\,2  & 17     & Y\\
28  &SSSPM J1102-3431       & -82.0\,$\pm$\,12.0 & -12.0\,$\pm$\,6.0 & 6& 70& 43\,$\pm$\,7  &+13.2& $\ldots$        & $\ldots$& Y\\
\enddata
\tablecomments{
Columns:
(1) TWA number, 
(2) other name,
(3) proper motion in RA (\mura\,$\equiv$\,$\mu_{\alpha}$\,cos\,$\delta$; 
ICRS frame; 1$\sigma$ errors),
(4) proper motion in Dec (\mudec; ICRS frame; 1$\sigma$ errors),
(5) proper motion reference,
(6) membership probability (\S\ref{member}),
(7) predicted distance from moving group method (\S\ref{distance})
and 1$\sigma$ uncertainty,
(8) predicted radial velocity from moving group method
(with uniform 1.6\,\kms 1$\sigma$ uncertainty),
(9) observed mean radial velocity,
(10) radial velocity reference,
(11) final TWA membership assessment.
For some binaries, I have estimated the systemic velocity by assuming a mass ratio.
These are probably good to $\sim$1\,\kms, and their errors have been marked with colons.
For the systemic RV of HD 98800, I adopted the component masses from \citet{Prato01}
and assumed M(Ab) = 0.5\,M$_{\odot}$. The radial velocity listed for TWA 15AB is that
measured for A, but the proper motion is for the photocenter of AB 
(TWA 15A = 2MASS J12342064-4815135, TWA 15B = 2MASS J12342047-4815195). Kinematic
distances to non-members (TWA 12, 17, 18, 19AB, and 24 are probably LCC members; TWA 22
may not be a member) should not be taken seriously. However, due to the similarity
in space motions between TWA and LCC, the distances to TWA 12,17,18,19AB, and 24
listed are probably within 5\% of the real distance if they are indeed LCC members. 
References: 
(1) \citet[][UCAC2]{Zacharias04},
(2) \citet[][SPM]{Platais98},
(3) calculated by the author using positions from the following catalogs:
GSC-ACT \citep{Lasker99}, 
USNO-A2.0 \citep{Monet98}, 
GSC 2.2 \citep{STSci01}, 
2MASS \citep{Cutri03}, 
DENIS \citep{DENIS03},
UCAC1 \citep{Zacharias01},
and \citet{Rousseau96},
(4) Tycho-2 \citep{Hog00},
(5) USNO-B1.0 \citep{Monet03},
(6) \citet{Scholz05},
(7) \citet{Torres03},
(8) \citet{delaReza89},
(9) \citet{Reid03},
(10) \citet{Torres95},
(11) \citet{Prato01},
(12) \citet{Barbier-Brossat00},
(13) \citet{Grenier99},
(14) \citet{Sterzik99},
(15) \citet[][as reported in \citet{Reid03}]{Webb99phd},
(16) \citet{Song03},
(17) \citet{Mohanty03}
}
\end{deluxetable}

\clearpage
\begin{deluxetable}{lcc}
\tabletypesize{\scriptsize}
\setlength{\tabcolsep}{0.03in}
\tablewidth{0pt}
\tablecaption{Trigonometric
vs. Cluster Parallaxes \label{tab:par}}
\tablehead{
{(1)}     &{(2)}                &{(3)}\\
{Name}    &{\parallax$_{trig}$} &{\parallax$_{clus}$}\\
{}        &{[mas]}              &{[mas]}\\
}
\startdata 
TW Hya    & 17.8$\pm$2.2 & 19.5\,$\pm$\,1.6\\
HD 98800  & 20.5$\pm$2.8 & 25.0\,$\pm$\,2.0\\
TWA 9     & 19.9$\pm$2.4 & 14.8\,$\pm$\,1.1\tablenotemark{a}\\
HR 4796   & 15.1$\pm$0.7 & 13.8\,$\pm$\,1.1\\
\enddata
\tablecomments{Columns: 
(1) common star name,
(2) weighted mean of Hipparcos and Tycho-1 trigonometric 
parallax \citep{Perryman97},
(3) parallax from cluster parallax method (this work).}
\tablenotetext{a}{Weighted mean of individual estimates
for TWA 9 A and B.}
\end{deluxetable}

\clearpage
\begin{deluxetable}{lcccc}
\tabletypesize{\scriptsize}
\setlength{\tabcolsep}{0.03in}
\tablewidth{0pt}
\tablecaption{Properties of 2M1207 A and B \label{tab:phot}}
\tablehead{
{(1)}                  &{(2)}               &{(3)}               &{(4)}  & {(5)} \\
{Property}             &{2M1207 A}          &{2M1207 B}          &{Units}& {Notes}}
\startdata 
Spectral Type          &  M8.5\,$\pm$\,1    & L7.25\,$\pm$\,2.25 &$\ldots$& 1\\
$K$                    & 11.96\,$\pm$\,0.03 & 16.93\,$\pm$\,0.11 & mag    & 2\\
$M_{K}$                &  8.32\,$\pm$\,0.27 & 13.30\,$\pm$\,0.29 & mag    & 3\\
$BC_K$                 &  3.12\,$\pm$\,0.14 &  3.25\,$\pm$\,0.14 & mag    & 4\\
log(L/L$_{\odot}$)     &--2.68\,$\pm$\,0.12 &--4.72\,$\pm$\,0.14 & dex    & 5\\
Mass \citep{Baraffe03} &  21(19-30)         & 3.3(2.3-4.8)       & \mjup  & 6\\
Mass \citep{Chabrier00}&  21(19-31)         & 3.2(2.3-4.8)       & \mjup  & 7\\
Mass \citep{Burrows97} &  20(17-24)         & 4.2(2.6-6.5)       & \mjup  & 8\\
\enddata
\tablecomments{
(1) spectral types from \citet{Chauvin04},
(2) $K$-band photometry from \citet{Chauvin04},
(3) absolute $K$ magnitudes assuming distance from \S\ref{distance}, and no reddening,
(4) bolometric corrections are from polynomial of \citet{Golimowski04}.
Error includes uncertainty in spectral type and rms of their  BC$_K$(SpT) fit, 
(5) luminosity using $M_{Ks}$ and $BC_K$ and assuming $M_{bol\odot}$ = 4.75,
(6) mass (and mass range) from COND evolutionary tracks of \citet{Baraffe03}.
(For all evolutionary tracks, the best interpolated mass estimate at age 8\,Myr is given,
followed by the extrema of the mass range considering the uncertainties 
in luminosity and age, where I assume an age of 8$^{+4}_{-3}$ Myr, following 
Chauvin et al. 2004),
(7) mass (and mass range) from DUSTY evolutionary tracks of \citet{Chabrier00},
(8) mass from evolutionary tracks of \citet{Burrows97}.
}
\end{deluxetable}

\end{document}